\documentclass[english,a4paper]{article}

\usepackage{graphicx}
\usepackage{amsmath}
\usepackage{amssymb}
\usepackage{mathtools}
\usepackage{color}

\begin{document}
\title{Simultaneous field-free molecular orientation and planar delocalization by THz laser pulses}

\author{D. Sugny\footnote{Laboratoire Interdisciplinaire Carnot de
Bourgogne (ICB), UMR 6303 CNRS-Universit\'e Bourgogne-Franche Comt\'e, 9 Av. A.
Savary, BP 47 870, F-21078 Dijon Cedex, France; dominique.sugny@u-bourgogne.fr}}

\maketitle

\begin{abstract}
This study shows the unexpected and counter-intuitive possibility of simultaneously orienting a molecule while delocalizing its molecular axis in a plane in field-free conditions. The corresponding quantum states are characterized and different control strategies using shaped THz laser pulses are proposed to reach such states at zero and nonzero temperatures. The robustness against temperature effects of a simple control procedure combining a laser and a THz pulse is shown. Such control strategies can be applied not only to linear molecules but also to symmetric top molecules.
\end{abstract}

\section{Introduction}
Quantum control tackles the question of bringing a quantum system from one state to another by means of external electromagnetic pulses. It has now become a well-recognized area of research with applications ranging from chemistry and quantum technologies to materials science and nanotechnology~\cite{Glaser:15,Brif:10,Koch:22}. In molecular physics, quantum control has been used in a large number of studies for manipulating molecular rotation, in particular to enhance molecular alignment and orientation in gas phase~(see \cite{Stapelfeldt:03,Seideman:05,Koch:19,Lemeshko:13,Prasad:21} and references therein). The control of such phenomena is key because they play a crucial role in a wide variety of domains such as chemical reaction dynamics, surface processing, catalysis and quantum computing to cite a few. In the case of a linear molecule driven by a linearly polarized laser field, alignment
means an increased probability distribution along the polarization axis whereas orientation requires in addition the same
or opposite direction as the polarization vector. Field-free alignment and orientation are investigated in this paper~\cite{Koch:19}. It is worth noting that for experiments requiring conditions without external electromagnetic fields, noticeable orientation and alignment that persists after the end of the pulse are of special importance. In this description, it seems obvious that any oriented state is also aligned while the converse is not true. Indeed, an aligned state can be defined as a quantum superposition of two oriented states along opposite directions. This study shows that this first statement is not entirely valid since there exist oriented states which can be (to some extent) also characterized by planar delocalization.

In a more general perspective, the control of molecular alignment is nowadays well understood in the adiabatic or sudden regime~\cite{Friedrich:95,Stapelfeldt:03,Seideman:05,Leibscher:03,Salomon:05,Leibscher:04} by means of laser pulses. In the past few years, different studies have extended the standard control framework by considering, e.g., the deflection of aligned molecules~\cite{Gershnabel:10}, collisional effects on quantum dynamics~\cite{Ramakrishna:05,Viellard:08,Viellard:13} or rotational echoes~\cite{Karras:15b}. The shaping of fied-free alignment dynamics has also been extensively investigated. A series of non-trivial extensions have been proposed. They go from planar alignment~\cite{Hoque:11}, unidirectional rotation of molecular axis~\cite{Korech:13,Steinitz:14,Karras:15}, alignment alternation~\cite{Daems:05a} to the control of rotational wave packet dynamics~\cite{Spanner:04}. The control of molecular orientation is not at the same stage of maturity even if, on the theoretical side, several control protocols to enhance the degree of orientation have been developed and applied with success~\cite{Averbukh:01,Coudert:17,Daems:05,Dion:05,Lapert:12,Sugny:14,
Tehini:08,Tehini:12,Henriksen:99,Kitano:11,Li:13,Omiste:16,Ortigoso:12,Wu:10,Yoshida:14,Spanner:12,
Shu:13,Liao:13,Znakovskaya:14,Kurosaki:14,Qin:14,Trippel:15,Kallush:15,Takei:16,Damari:17,Hong21}. Some of them have been implemented
experimentally~\cite{Fleischer:11,Babilotte:16,Ghafur:09,Goban:08,Znakovskaya:09,Frumker:12,Frumker:12b,Dimitrovski:11}, in particular in field-free conditions by using TeraHertz (THz) laser fields~\cite{Fleischer:11,Babilotte:16}. It is worth noting that most of the control strategies developed so far have only investigated the optimization of the degree of orientation without generating other dynamics~\cite{Tehini:19}.

This paper explores another aspect of field-free molecular orientation, that is the unexpected and counter-intuitive possibility to orient a molecule while delocalizing its molecular axis in a plane orthogonal to the orientation direction. Introducing a new observable combining the degrees of orientation and alignment of the system, the corresponding quantum states are identified as specific eigenvectors of this operator. They are then completely characterized both from a classical and a quantum point of view. The physical limits to this simultaneous dynamic are also established. Using optimal control theory (OCT)~\cite{Glaser:15,Koch:22}, a THz field which brings the system to the target
state is designed at zero temperature. Such optimal solutions are of remarkable efficiency, close to 100\%. Numerical results are presented for the CO molecule. Despite its relative complexity, the control procedure could be, in principle, experimentally implemented by standard pulse shaping techniques. A less demanding quantum state which is oriented but not aligned is considered at nonzero temperature. A  simple bipulse process is then proposed to generate these states. It is composed of a short laser pulse followed by a THz half-cycle pulse (HCP)~\cite{Koch:19} after a quarter rotational period. It is shown that this control strategy is efficient, robust to temperature effects and experimentally easier to implement. This control procedure is applied to the linear CO molecule but also to a symmetric top molecule, CH$_3$I.

The paper is organized as follows. The quantum states leading simultaneously to orientation and planar delocalization are characterized and described in Sec.~\ref{sec2}. Different control strategies to reach these states are proposed in Sec.~\ref{sec3} both at zero and nonzero temperatures. Conclusions and prospective views are given in Sec.~\ref{sec4}.
\section{Description of the classical and quantum states}\label{sec2}
The time evolution of the rotation of a polar linear molecule subjected to a linearly polarized THz pulse of amplitude $E(t)$ at zero temperature is governed by the following Schr\"odinger equation written in the rigid rotor approximation~\cite{Koch:19}
$$
i\frac{d}{dt}|\psi(t)\rangle =H|\psi(t)\rangle
$$
with $|\psi(t)\rangle$ the state of the system and the Hamiltonian $H$ given by
$$
H=BJ^2-\mu_0\cos\theta E(t)
$$
where $B$ and $\mu_0$ are respectively the rotational constant and the permanent dipole moment. $J^2$ is the angular momentum operator and $\theta$ the polar angle between the molecular axis and the field polarization direction corresponding to the $z$- axis of the laboratory frame $(x,y,z)$. The eigenvectors of $J^2$ in this frame are denoted $|jm\rangle$ with $j\geq 0$ and $-j\leq m\leq j$, the $z$- axis being the quantum axis used to define these rotational states. Note that the units used throughout this paper are atomic units unless otherwise specified. The degrees of orientation and alignment are respectively measured by $\langle\cos\theta\rangle$ and $\langle\cos^2\theta\rangle$. In field-free conditions, the two expectation values oscillate with a period $\pi/B$ corresponding to the rotational period of the molecule.
A planar delocalization is achieved when $\langle\cos^2\theta\rangle <\frac{1}{3}$ (1/3 being the degree
of alignment at thermal equilibrium) while the molecule is said to be oriented along the $z$- axis if $\langle\cos\theta\rangle\simeq \pm 1$, or at least if $\langle\cos\theta\rangle\neq 0$.

A field-free simultaneous orientation and planar delocalization combines at the same time the degrees of orientation and alignment. The corresponding observable therefore both depends on $\cos\theta$ and $\cos^2\theta$. This choice is not unique but a convenient one is the observable $F$ defined as
$$
F=\cos\theta-a\cos^2\theta
$$
where $a$ is a positive parameter which expresses the relative weight
between the orientation and the alignment of the state. At this point, it is clear that the targeted quantum state must maximize $\langle F\rangle$.
\begin{figure}[tb]
  \centering
  \includegraphics[width=1.0\linewidth]{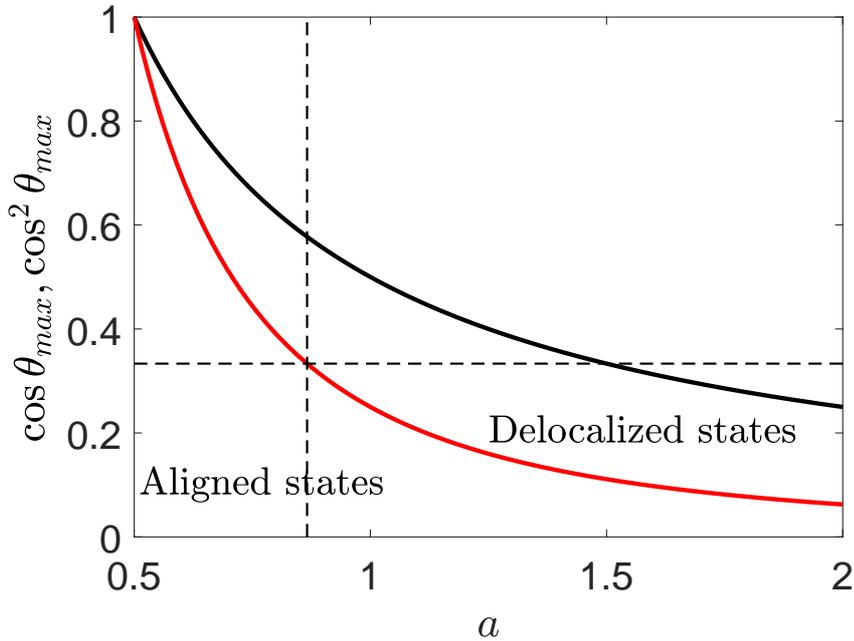}
  \caption{Evolution of $\cos(\theta_{max})$ (black line) and $\cos^2(\theta_{max})$ (red line) as a function of the parameter $a$. $\theta_{max}$ is the angle which maximizes the figure of merit $F$ for a given value of $a$. The horizontal dashed line delimits the region of aligned and delocalized states. The area to the right of the vertical dashed line (of equation $a=\frac{\sqrt{3}}{2}$) corresponds to simultaneous orientation and planar delocalization}
  \label{fig1}
\end{figure}
A first analysis of the behavior of $F$ can be done in the classical limit. In this case, the maximum of $F$ occurs for $\theta_{max}=\textrm{acos}(\frac{1}{2a})$ leading to $\cos\theta_{max}=\frac{1}{2a}$. Figure~\ref{fig1} displays the evolution of the two classical observables with respect to $a$. It can be deduced that the value of $a$ which maximizes $F$ is $a=\frac{1}{2}$ for $\theta_{max}=0$ and $\cos\theta_{max}=\cos^2\theta_{max}=1$. In this case, a simultaneous molecular orientation and alignment is achieved. Isotropic distribution of the probability density is obtained for $\cos^2\theta=\frac{1}{3}$, which gives $a=\frac{\sqrt{3}}{2}$ and $\cos\theta=\frac{1}{\sqrt{3}}$, as indicated in Fig.~\ref{fig1}. More interesting, when $a>\frac{\sqrt{3}}{2}$, a planar delocalization for an oriented state is produced with $\cos^2\theta<\frac{1}{3}$ and $\cos\theta>0$.    The results can be summarized as follows:
\begin{itemize}
\item For $a\in [\frac{1}{2},\frac{\sqrt{3}}{2}]$, the molecule is aligned and oriented with $\cos^2\theta\geq 1/3$ and $\cos\theta>0$.
\item For $a\geq \frac{\sqrt{3}}{2}$, the molecule is delocalized and oriented with $\cos^2\theta\leq 1/3$ and $\cos\theta>0$.
\end{itemize}
It is shown below that such classical limits give a good estimate of the expectation values of the observables that can be achieved in the quantum regime.

The next step is to describe this phenomenon in the quantum setting. A major mathematical difficulty of this description comes from the infinite-dimensional Hilbert space of the rotational system. This problem is circumvented by considering a reduction of the original Hilbert space to a finite-dimensional one. From a physical point of view, this reduction can be justified by the fact that the system is only subjected to a finite number of applied pulses with moderate amplitude. Since such pulses transfer finite amounts of energy to the system, this latter stays thus essentially confined in a finite-dimensional subspace. This reduction considerably simplifies the definition of the target state and the analysis of its properties. Note that the strategy used in this paper emerges as a specific example of a more general analysis already applied to maximize, e.g., molecular orientation and alignment~\cite{Sugny:04,Sugny:05,Lapert:09,Lapert:11}. A natural reduction of the Hilbert space consists in introducing a finite Hilbert space $\mathcal{H}_{j_{max}}$ spanned by the lowest rotational states $\{|j,m\rangle, j\leq j_{max}\}$. The idea is then to identify the target state $|\psi_T\rangle$ in $\mathcal{H}_{j_{max}}$, as the state maximizing the simultaneous orientation and planar delocalization in this subspace, i.e. maximizing $\langle \psi_T|P_{j_{max}}FP_{j_{max}}|\psi_T\rangle$, where $P_{j_{max}}$ is the projector onto the space $\mathcal{H}_{j_{max}}$. It is then straightforward to show that this state corresponds to the eigenvector of $P_{j_{max}}FP_{j_{max}}$ with the highest eigenvalue~\cite{Sugny:04}. Note that this method can be generalized to density matrices at nonzero temperatures~\cite{Sugny:05}. However, the generalization at nonzero temperature is not straightforward because, in this case, the state of the system is described by a density matrix (see Sec.~\ref{sec3} for technical details). The idea would then be to find the density operator which maximizes the expectation value of the observable $F$, while being linked to the initial state by unitary evolution. This leads to a rather involved computation which is beyond the scope of this manuscript.

\begin{figure}[tb]
  \centering
  \includegraphics[width=1.0\linewidth]{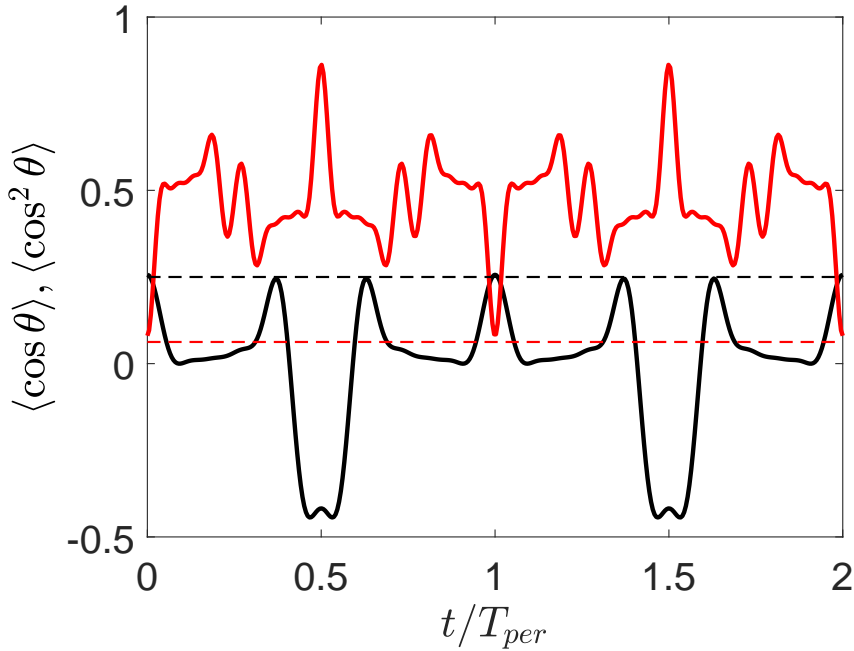}
  \caption{Contour plot of the maximum of $\langle\cos\theta\rangle$ (top panel) and $\langle\cos^2\theta\rangle$ (bottom panel) as a function of $a$ and $j_{max}$}
  \label{fig5}
\end{figure}

This general discussion is illustrated by a series of numerical results on a fictive molecule with $B=1$. Figure~\ref{fig5} displays the maximum of $\langle\cos\theta\rangle$ and $\langle\cos^2\theta\rangle$ with respect to the two parameters $j_{max}$ and $a$. Two different zones can be clearly distinguished. A region of simultaneous high degrees of orientation and alignment can be found in the top left part of the two panels. This zone corresponds typically to small values of $a$, lower than 1. More unexpected, a flat zone for $a\geq 2$ and $j_{max}\geq 7$ also appears in the right parts of Fig.~\ref{fig5}. In this case, $\langle\cos^2\theta\rangle <\frac{1}{3}$ so the molecule is delocalized in the $(x,y)$- plane, but with a noticeable orientation larger than 0.2.

\begin{figure}[tb]
  \centering
  \includegraphics[width=1.0\linewidth]{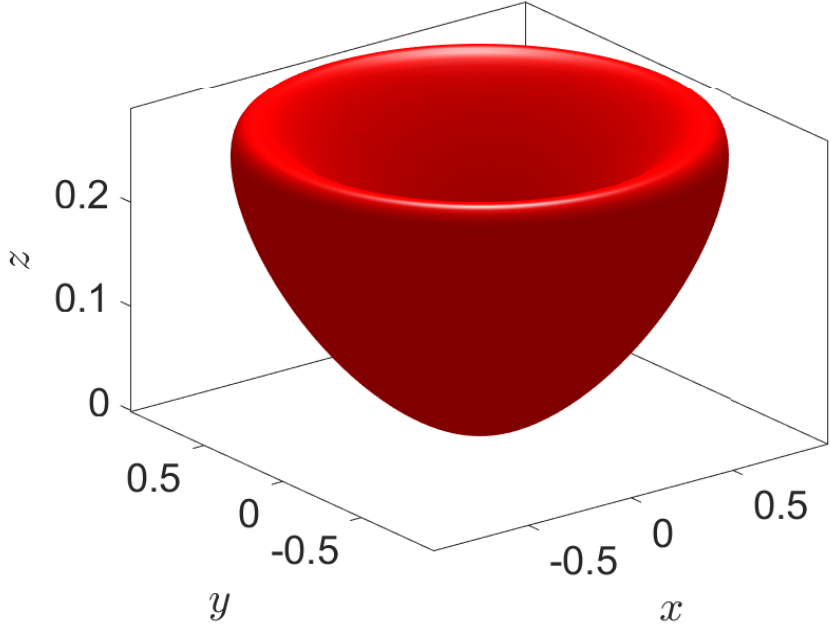}
  \caption{Probability density of the quantum state $|\psi_T\rangle$ maximizing simultaneously the orientation and the planar delocalization for $a=2$ and $j_{max}=10$}
  \label{fig4}
\end{figure}

More insight about this unusual behavior is given in Fig.~\ref{fig4} with the probability density of $|\psi_T\rangle$. This density has the shape of a bowl without handle and with a rotational symmetry around the $z$- axis. It clearly corresponds to an oriented state since the probability density belongs to the subspace with $z\geq 0$. However, the flared shape of the bowl allows at the same time to have a noticeable planar delocalization in the $(x,y)$- plane. This density is also very different from those obtained with oriented and aligned states that have a cigar shape along the $z$- axis.
Note that there exists a target state for every value of $j_{max}$, but with very similar properties when $j_{max}\geq 7$. The state that is generally reached by a control process features a quantum superposition of different target states.

\begin{figure}[tb]
  \centering
  \includegraphics[width=1.0\linewidth]{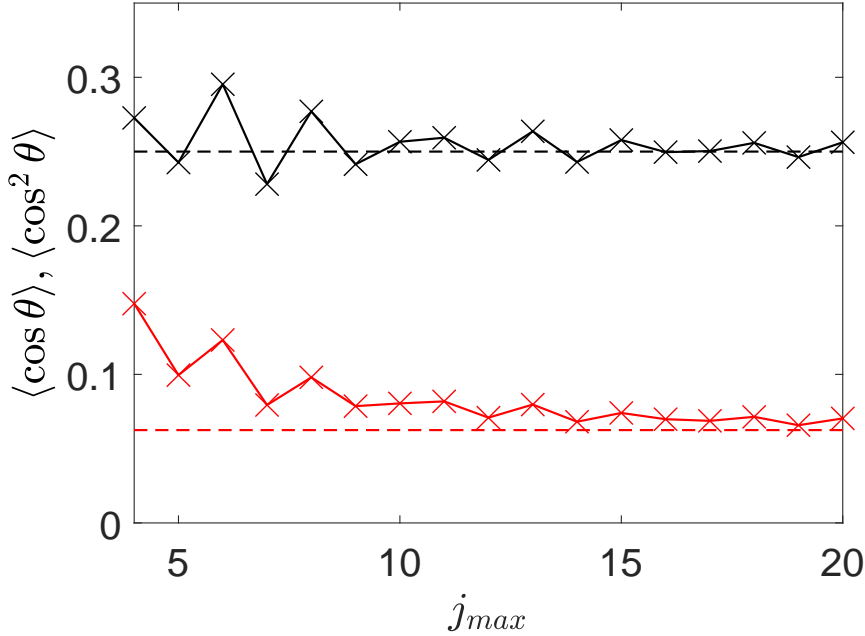}
  \caption{Evolution of the expectation values $\langle\cos\theta\rangle$ (black line) and $\langle\cos^2\theta\rangle$ (red line) in the state $|\psi_T\rangle$ of a fictive molecule as a function of $j_{max}$ for $a=2$ (crosses). The solid lines are just to guide the lecture. The horizontal dashed lines represent the classical values of $\cos\theta$ and $\cos^2\theta$ for $\theta_{max}=\frac{1}{2a}$}
  \label{fig3}
\end{figure}

Another interesting characteristic of $|\psi_T\rangle$ can be found in Fig.~\ref{fig3} in which the degrees of orientation and alignment are plotted as a function of $j_{max}$ for a fixed value of $a$. It can be seen that for $j_{max}\geq 10$, the two quantum expectation values are almost constant and very close to their classical estimates given in Fig.~\ref{fig1}. This result confirms the fact that this state depends little on the value of $j_{max}$ and that useful information can be gained from a classical analysis of the rotational dynamics. As already mentioned, the choice of the parameter $a$ allows to weight the relative importance of orientation and planar delocalization in the target state.

\begin{figure}[tb]
  \centering
  \includegraphics[width=1.0\linewidth]{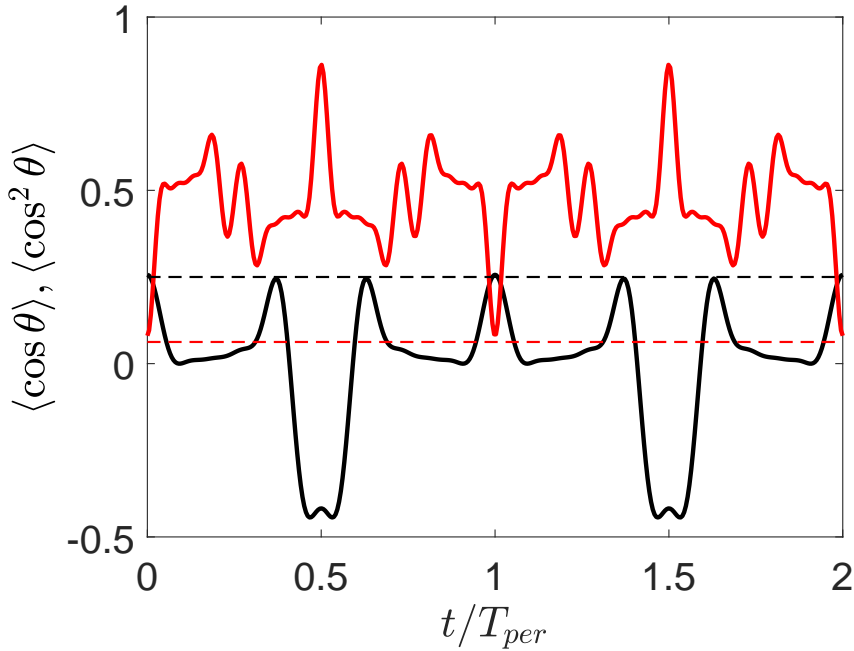}
  \caption{Field-free time evolution of the expectation values $\langle\cos\theta\rangle$ (black line) and $\langle\cos^2\theta\rangle$ (red line) of a fictive molecule at $T=0$~K. The initial state at $t=0$ is $|\psi_T\rangle$. The parameters $a$ and $j_{max}$ are set to 2 and 10. The horizontal dashed lines represent the classical values of $\cos\theta$ and $\cos^2\theta$ for $\theta_{max}=\frac{1}{2a}$}
  \label{fig2}
\end{figure}

Figure~\ref{fig2} gives a view of the rotational dynamics starting from $|\psi_T\rangle$. Different revivals of simultaneous orientation and delocalization are observed at multiple times of $T_{per}$. Note that it is only around such times that $\langle\cos^2\theta\rangle<\frac{1}{3}$. Here again, the quantum expectation values are very close to their classical counterparts. At revival times $t=(2n+1)\frac{T_{per}}{2}$ for integers $n$, the dynamic goes through a state which is aligned and oriented.

\section{Control of simultaneous orientation and planar delocalization at zero and nonzero
temperatures}\label{sec3}
\begin{figure}[tb]
  \centering
  \includegraphics[width=1.0\linewidth]{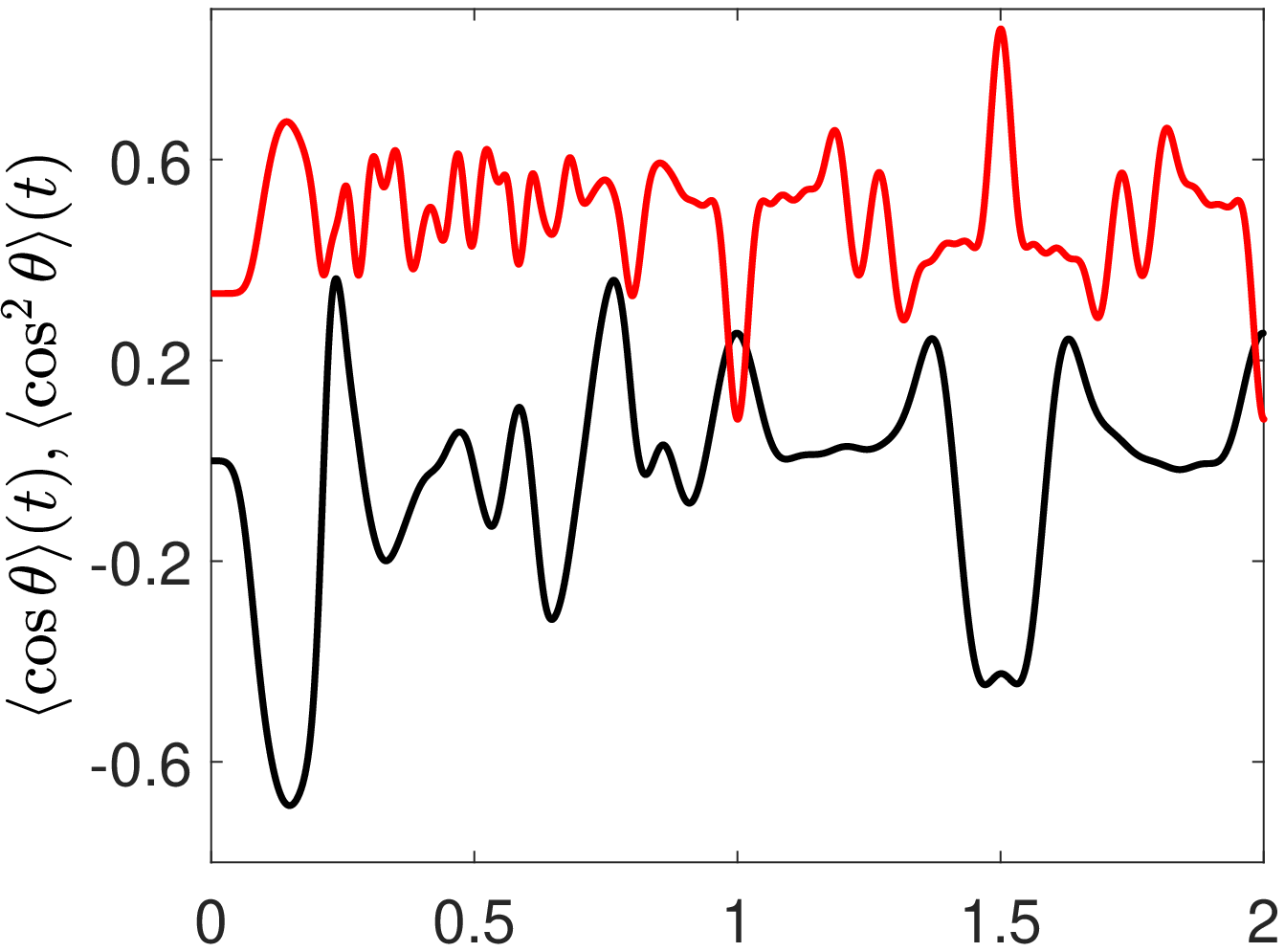}
  \includegraphics[width=1.0\linewidth]{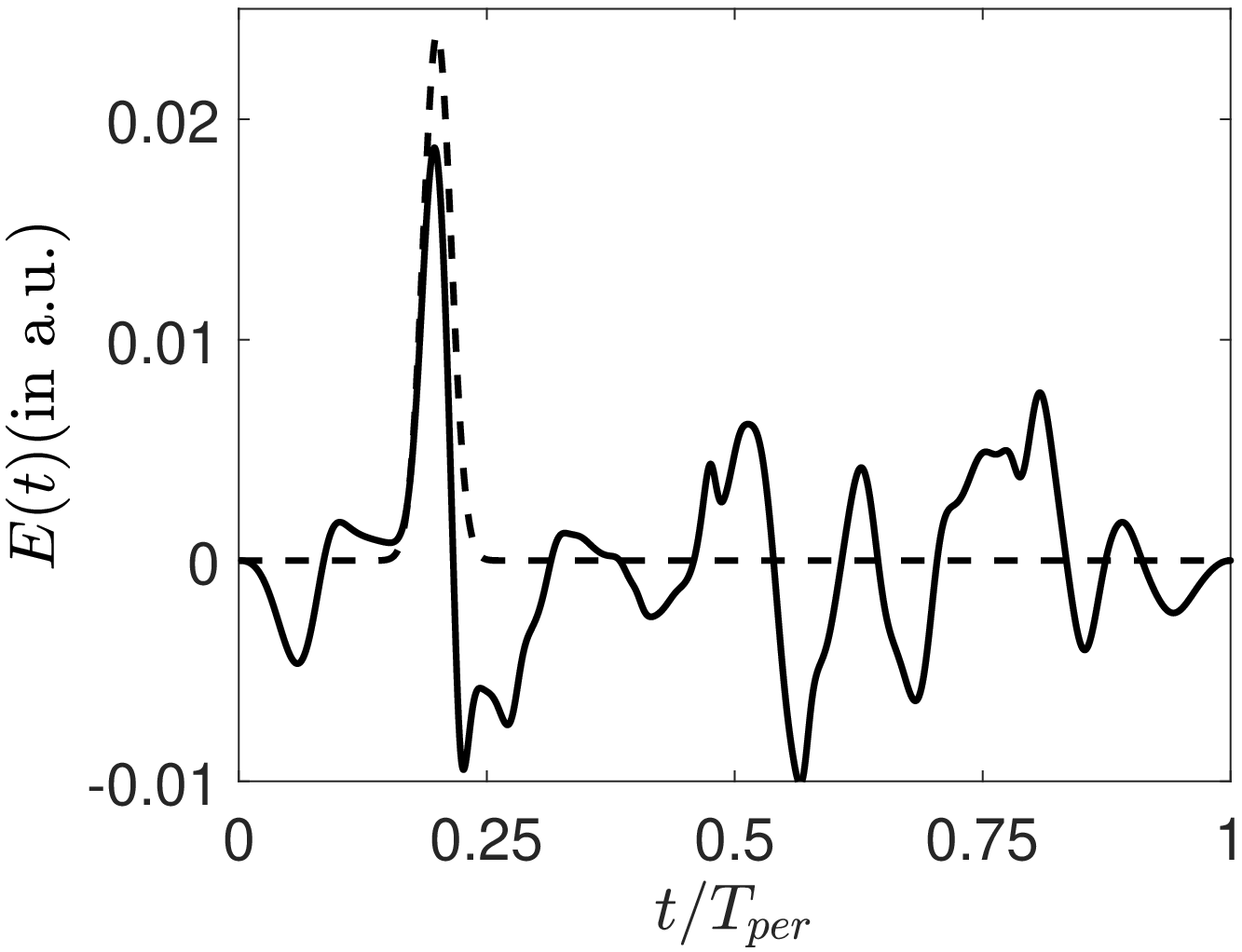}
  \caption{(Top) Time evolution of $\langle\cos\theta\rangle$ (black line) and $\langle\cos^2\theta\rangle $ (red line) for the CO molecule at $T=0$~K under the action of the optimized pulse (bottom) followed by a field-free evolution of one rotational period. Numerical parameters are set to $a=2$ and $j_{max}=10$}
  \label{fig6}
\end{figure}

Having defined the target states, the next step consists in designing control strategies to
reach a given target state or a superposition of such states. Control fields using OCT~\cite{Glaser:15,Brif:10,Boscain:21} are first computed. A standard gradient-based iterative algorithm whose efficiency has been shown in a variety of studies for rotational dynamics~\cite{Salomon:05,Brif:10,Coudert:17} is applied to the CO molecule. Molecular  parameters are taken to be $B=1.9313$~cm$^{-1}$ and $\mu_0=0.112$~D. As an illustrative example, a pulse duration of one rotational period and a target
state corresponding to $j_{max}=10$ and $a=2$ are considered. Similar results have been obtained for other values of $j_{max}$ and $a$. The goal of the control process is to maximize the projection of the final state onto the target. A Gaussian pulse centered at $t_0 = T_{per}/5$ is taken as a guess field for the optimization procedure with $E(t) = E_0 e^{\frac{(t-t_0)^2}{2\sigma^2}}$ where $I_0 = 20$~TW/cm$^2$ is the maximum intensity of the electric field and $\sigma = \pi/(50 B)$. A final projection larger than 0.99 has been achieved. Figure~\ref{fig6} displays the results of the optimization procedure with the optimal solution computed by the algorithm. A simultaneous orientation and planar delocalization is obtained at $t=T_{per}$ when the control field is switched off.

\begin{figure}[tb]
  \centering
  \includegraphics[width=1.0\linewidth]{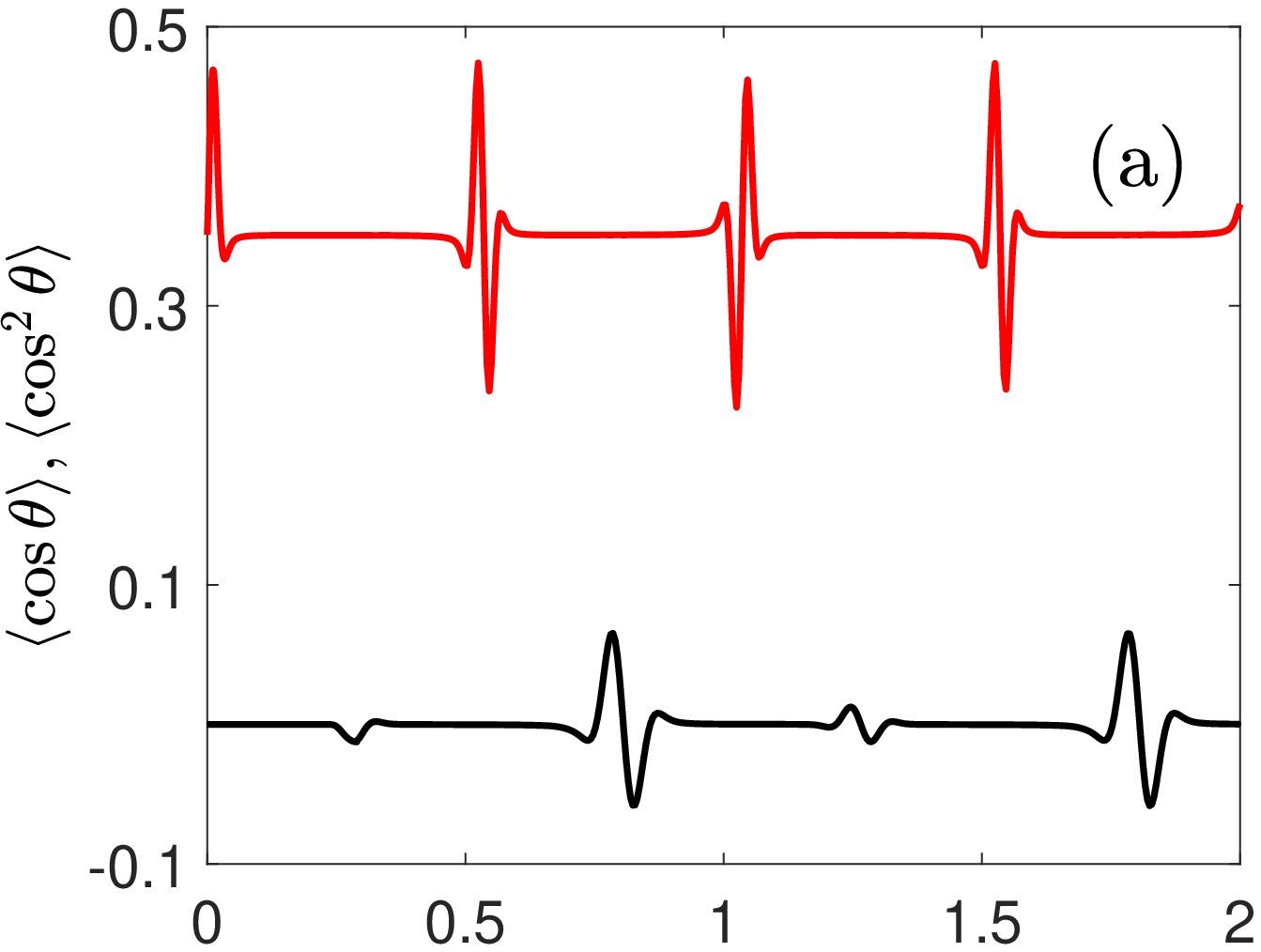}
  \includegraphics[width=1.0\linewidth]{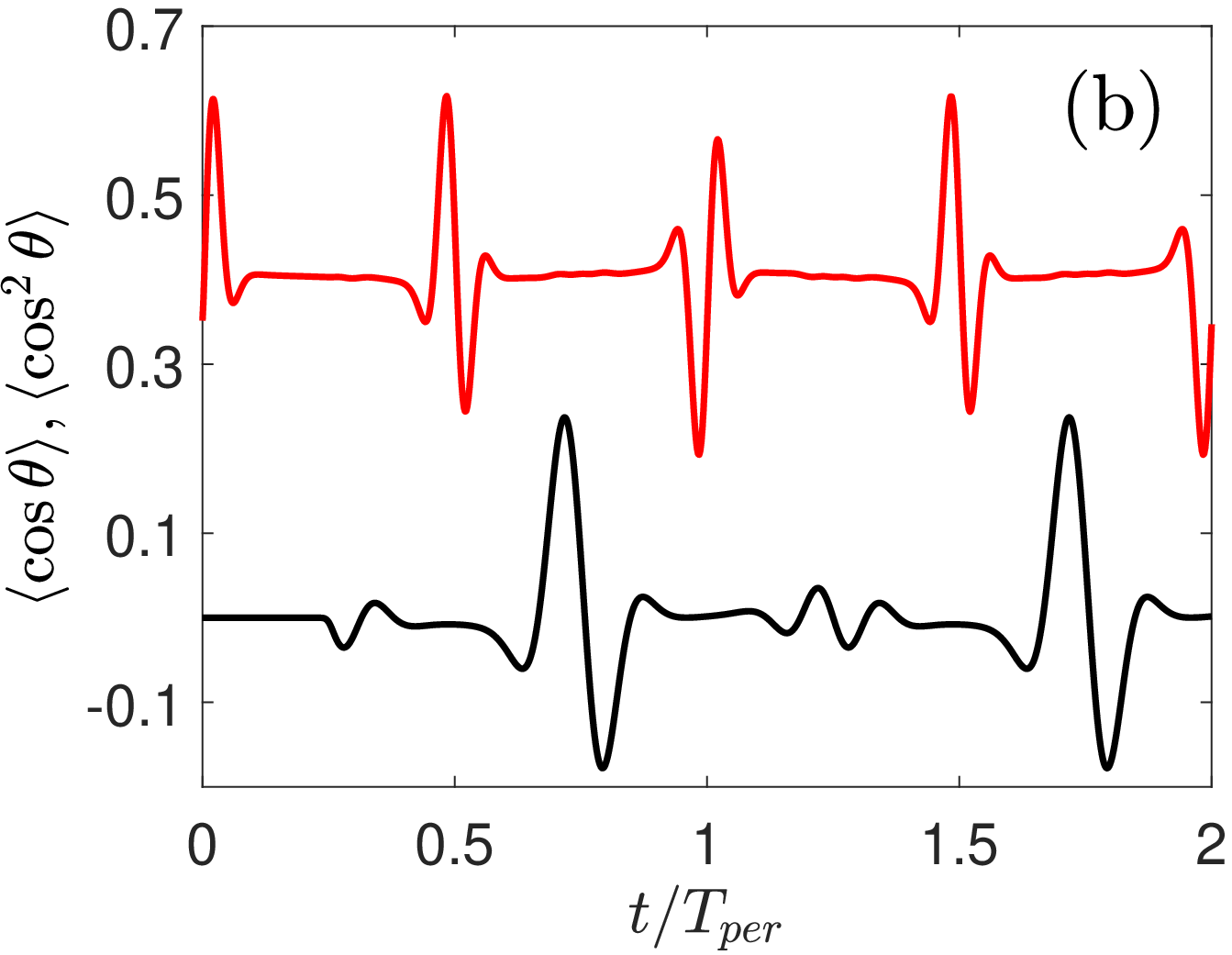}
  \caption{Time evolution of $\langle\cos\theta\rangle$ (black line) and $\langle\cos^2\theta\rangle $ (red line) for the CO molecule at $T=200$ (panel (a)) and 30~K (panel (b)) generated by a laser pulse followed by a HCP}
  \label{fig7}
\end{figure}

\begin{figure}[tb]
  \centering
  \includegraphics[width=1.0\linewidth]{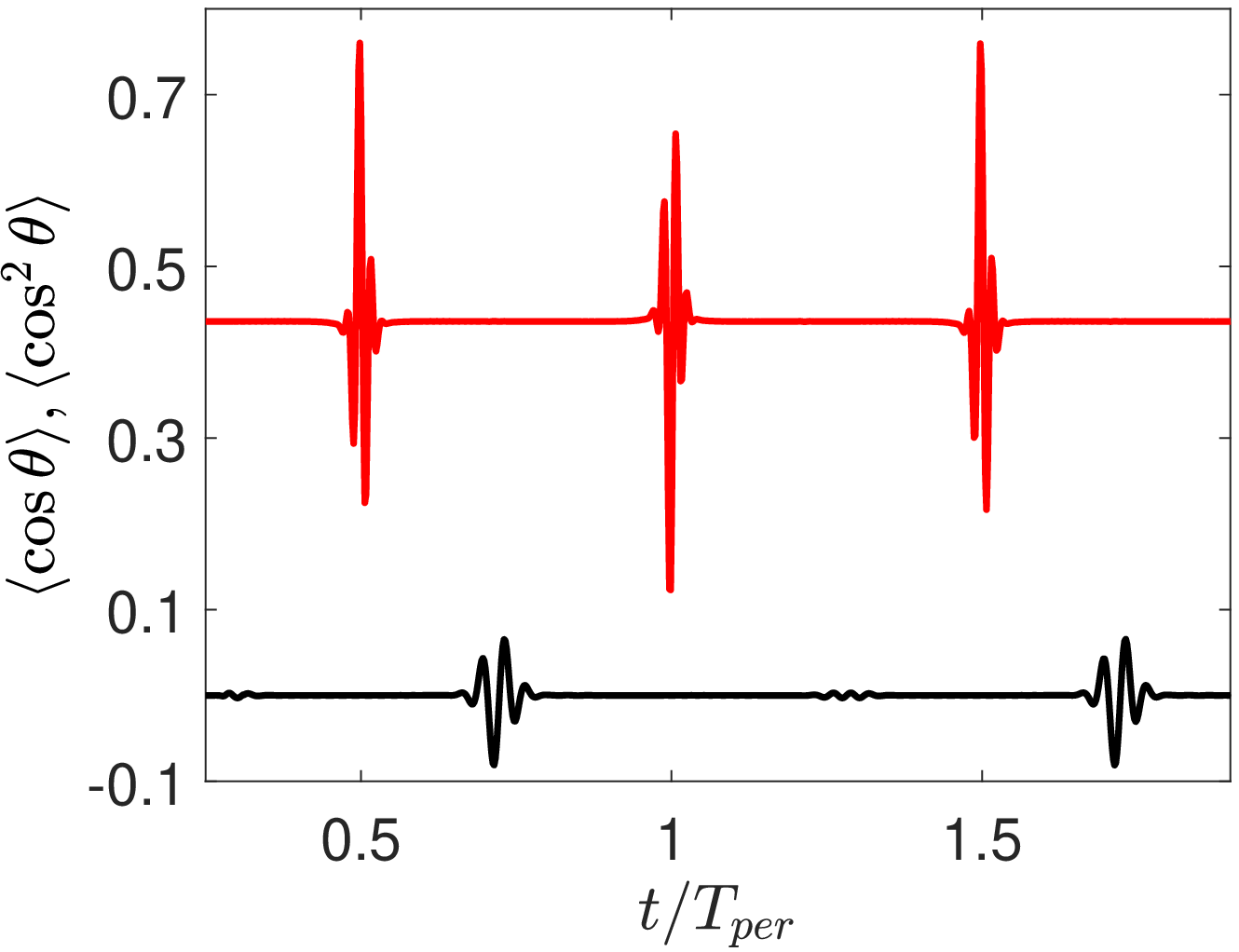}
  \caption{Same as Fig.~\ref{fig7} but for the CH$_3$I molecule at $T=30$~K. The HCP is replaced by a single-cycle pulse in the control process. Note that the range of time starts at time $t/T_{per}=0.25$ in order to highlight the field-free evolution}
  \label{fig8}
\end{figure}

In view of experimental applications, it is also interesting to find simpler strategies composed of a series of short pulses. Even if these control processes have an efficiency which is less than that of optimal control, they also have the decisive advantage of being robust with respect to experimental constraints such as temperature effects. A control procedure at nonzero temperatures is now investigated. In this case, the quantum system is described by a density operator $\rho(t)$ whose dynamics are governed by the von Neumann
equation~\cite{Koch:19}
$$
i\frac{d\rho}{dt}=[H(t),\rho].
$$
The initial condition at $t=0$, i.e. when the laser is switched on, is the Boltzmann distribution
$$
\rho(0)=\frac{1}{Z}\sum_{j=0}^\infty\sum_{m=-j}^je^{-Bj(j+1)/(k_BT)}|jm\rangle\langle jm|
$$
where $Z=\sum_{j=0}^\infty\sum_{m=-j}^je^{-Bj(j+1)/(k_BT)}$ is the partition function, $k_B$ the Boltzmann constant and $T$ the temperature of the sample. Here, the degree of orientation is given by $\langle\cos\theta\rangle =\textrm{Tr}[\rho\cos\theta]$. At nonzero temperature, it is difficult to reach states that are both oriented and with a planar delocalization. Note that the target states defined at zero temperature in Sec.~\ref{sec2} could be generalized to the density matrix formalism using the results described in Ref.~\cite{Sugny:05}. However, the computation is rather involved and leads to complex pulses. To limit the complexity of the control processes, the idea is thus to relax these conditions and to consider less demanding states which are oriented but not   aligned, i.e. with $\langle\cos^2\theta\rangle\simeq 1/3$. A simple control strategy can be used to this aim. It is based on the application of a short laser pulse followed by a HCP, the two pulses are temporally delayed by $T_{per}/4$. The two pulses are gaussian and linearly polarized along the $z$- axis. Note that similar control procedures have been proposed to enhance molecular orientation~\cite{Gershnabel:06,Gershnabel:06b,Daems:05}. Numerical results are displayed in Fig.~\ref{fig7} for the CO molecule at two different temperatures. The intensity of the laser is  50~TW/cm$^2$ with a duration (FWHM) of 50~fs while the HCP has a duration of 100 fs, the corresponding electric field having a peak amplitude of $10^8$~V/m. It can be seen in Fig.~\ref{fig7} that the first laser field at time $t=0$ produces a non adiabatic alignment featured by revivals occurring at fractional times of the rotational period (multiples of $T_{per}/2$). The second pulse breaks the symmetry of the dynamics and generates oriented states. The collective action of the two pulses results in a transient orientation at times $T_{per}/4+nT_{per}$, for integers $n$, with an alignment equal to the permanent alignment of the molecule. The dynamic alternates between an aligned state with no orientation and an oriented state with only a permanent alignment. Note that the control process is robust with respect to temperature as shown in Fig.~\ref{fig7} in which the same control procedure has been used at two different temperatures. This phenomenon can be enhanced to some extent by using a laser pulse followed by a train of HCPs applied at times $T_{per}/4+nT_{per}/2$ with adjusted intensity~\cite{Leibscher:03,Sugny:04}.
The control strategy extends to symmetric top
molecules as illustrated in Fig.~\ref{fig8} for the CH$_3$I molecule at $T=30$~K. In the control process, the intensity of the laser is set to 50~TW/cm$^2$ and the HCP is replaced by a single-cyle THz pulse of peak amplitude $2\times 10^7$~V/m. This latter pulse has been recently used experimentally to orient this molecule (see~\cite{Babilotte:16} for the molecular parameters and technical details). Here again, a noticeable orientation is achieved with only a permanent alignment. The robustness against temperature effects and variations of pulse amplitude has been also verified.

\section{Conclusion}\label{sec4}
This study has investigated the extent to which
a molecule can be simultaneously oriented and delocalized in a plane in field-free conditions. After a classical description of the phenomenon, the corresponding quantum states have been described in details at zero temperature. Such target states can be reached with a very good efficiency by using optimal control procedures, the price to pay being a relatively complex control pulse.  The efficiency of this method is only limited by the maximum field intensity that can be used to prevent molecular ionization. At nonzero temperature, a simpler control strategy has been proposed allowing to generate less demanding dynamics. This procedure can be used for linear molecules but also symmetric top molecules. Such a phenomenon is interesting from a fundamental point of view and it allows to show that unexpected results can be obtained in the quantum regime by shaping at will the probability density of the rotational states~\cite{Tehini:19}. Finally, it should be possible to reach such states by using a spectrally
shaped two-color laser pulse~\cite{Tehini:08,Znakovskaya:09}. This issue which
goes beyond the scope of this study is an interesting generalization of the results presented in this paper.\\

\noindent\textbf{Acknowledgment.} I thank Dr. M. Lapert and M. Ndong for many helpful discussions and for providing me the idea at the origin of this paper.

\bibliographystyle{unsrt}

\end{document}